\documentclass{macro}    

\newdisplay{guess}{Conjecture}

\input psfig.tex
\begin{document}                                                                                   
\input psfig.tex
\begin{article}
\begin{opening}         
\title{On the Formation of Accretion-powered Galactic and 
Extra-galactic Outflows}
\author{Tapas K. \surname{Das}}  
\runningauthor{Tapas K. Das}
\runningtitle{Accretion-powered Galactic and Extra-galactic Jets}
\institute{IUCAA Post Bag 4 Ganeshkhind Pune 411 007 INDIA\\
Email: tapas@iucaa.ernet.in}
\begin{abstract}
Though widely observed to be emanating from a
variety of astrophysical sources, the underlying physical
mechanism behind the formation of galactic and extragalactic
outflows is still enshrouded in a veil of mystery.  In
addition, it has not been possible to calculate accurately the
amount of matter expelled in these events.  In this article we
present a {\em non-self-similar analytical model}, which, for
the first time, we believe is able to explain the outflow
formation phenomenon as well as compute the mass outflow rate
by simultaneously solving the equations governing the exact
transonic accretion and outflow.  Our model predicts the
dependence of this rate on various flow parameters as well as
the {\em exact} location from where the outflows are launched.
\end{abstract}
\keywords{AGN-quasars - microquasars - jets - accretion, accretion discs -
black hole physics - hydrodynamics - outflow - wind - shock-wave}
\end{opening}           
\section {Introduction}
It is now an well established fact that Quasars and Microquasars suffer 
massloss through outflows and jets 
(Mirabel \& Rodriguez, 1999, Ferrari, 1998, Begelman, Blandford \& Rees,
1984). These galactic and extra-galactic
jet sources are commonly believed to harbour accreting compact objects
at their hearts as the prime movers for almost all non stellar energetic
activities around them including the production of bipolar outflows and
relativisic jets. Unlike normal stellar bodies, compact objetcs do not have
their own physical atmosphere from where matter could be ripped off as winds,
hence outflows from the viscinity of these prime movers {\it 
have to be generated 
only from the accreting material}. So instead of separately 
investigating the jets and accretion processes as two disjoint issues around
the dynamical centre of the galactic and extra-galactic jet sources, it is 
absolutely necessary to study these two phenomena within the same framework and 
any consistent theoretical model for jet production should explore the outflow 
formation only from the knowledge of accretion parameters. 
Also to be noted that while self-similar
models are a valuable first step, they can  never be the full answer,
and indeed any model which works equally well at all radii is fairly
unsatisfactory to prove its viability. Thus the preferred model
for jet formation must be one which is able to
select the {\it specific} region of jet formation. 
Keeping these basic facts at the back of our mind, we propose a {\it non
self-similar analytical model} capable of self-consistently exploring the
hydrodynamic origin of accretion powered jets/outflows emanating 
out from galactic and extra-galactic sources. Using this model it 
has been possible to simultaneously solve the accretion and wind equations
to compute (from the first principle) what fraction of accreting material 
is being blown as wind and the exact location (distance measured from the
central accretor) of the jet launching zone has been successfully pointed
out. \\
Denoting ${\dot M}_{in}$ and ${\dot M}_{out}$ to be the instantaneous 
time rate of inflow and outflow around an accreting compact object, we define 
a quantity ``Mass Outflow Rate ($R_{\dot M}$)" as the ratio of  ${\dot M}_{out}$
to  ${\dot M}_{in}$, which is basically a measure of the fraction of barionic
accreting material being expelled as outflows/jets. Our major aim was to 
compute the exact value of $R_{\dot M}$ in terms of {\it minimum} number of
accretion parameters and to investigate the dependence of $R_{\dot M}$ on 
various flow parameters. We do our calculation for accretion with 
considerable intrinsic angular momentum (Disc-Outflow system, see \S 2 for
detail) and for spherical/quasi-spherical accretion with negligibly small 
angular momentum (see \S 3 for detail).\\
Because of the absence of any matter donating real physical atmosphere
around accreting compact objects, we first need to incorporate some virtual surfaces
around these objects from where outflows may be generated. Formation of these
surfaces are explained in \S 2.1 (for the disc-outflow system) and in \S 3.1
(for spherical/quasispherical accretion-outflow system).\\
In this article, we are basically interested in highlighting
the conceptual ideas behind
our model rather than presenting the mathematical details, thus we keep this
article free from any formulae or equations. Interested readers may kindly 
refer respective works (Das, 1998, 1999, 1999a, ,1999b, 2000, 2000a, Das \&
Chakrabarti, 1999 (DC99 hereafter)) for detail mathematical formalism.
\section{The Disc-Outflow System}
\subsection{Model Description and Prescription for the 
Formation of Outflow Generating Surface}
We consider thin, axisymmetric polytropic inflows around a Scwarzschild
Black hole in vertical equilibrium.  We
ignore the self-gravity of the flow and calculations are done 
using Paczy\'
nski-Wiita (Paczyn\'ski \& Wiita, 1980)
potential which mimics surroundings of the Schwarzschild black hole.
Considering the inflow to be polytropic, we explore
both the polytropic and the isothermal outflow.\\
Due to the fact that close to the BH the radial component of the 
infall velocity of accreting material would be enormously high, 
viscous time scale would be much longer than the infall time scale and
a rotating inflow 
entering into a black hole  will have almost constant 
specific angular momentum
close to the black hole for any moderate viscous stress (Das, 1998 and 
DC99).
Though at the outer edge of the accretion disk
the angular momentum distribution
may be Keplerian or even super-Keplerian, matter would be highly
sub-Keplerian close to the
black hole to satisfy the inner boundary condition at the event horizon of 
the BH
(see Chakrabarti, This volume and references therein).
This almost constant angular momentum
produces a very strong centrifugal force  which
increases much faster compared to the gravitational force and becomes comparable at
some specific radial distance, location of which is easy to compute.
Here, (actually, a little farther out, due to
thermal pressure) matter starts
piling up and produces the centrifugal pressure supported boundary layer
(CENBOL). Further close to the black hole, the gravity always wins
and matter enters the horizon supersonically after passing
through a sonic point. 
Formation of CENBOL may be attributed to the shock formation in accreting fluid 
or to the maximisation of polytropic pressure of the inflow.
In CENBOL region
the flow becomes hotter and denser
and for all practical purposes
behaves as the stellar atmosphere so far as the formation of
outflows are concerned.
A part of the hot and dense shock-compressed inflowing material
is then `squirt' as outflow from the CENBOL.
In case where the shock does
not form,
regions around pressure maximum achieved just outside the inner sonic
point of the {\it inflow} would also drive the flow outwards. 
The outflow is shown to be thermally and
centrifugally accelerated but is assumed to be 
confined by external pressure of the
ambient medium.  Subsonic outflows originating
from CENBOL would pass through sonic points and reach far distances
as in wind solution.\\
It is interesting to `visualize' how the combined accretion-outflow
system along with the central accretor would `look like' in reality.
\begin{figure}
\vbox{
\vskip -0.0cm
\centerline{
\psfig{figure=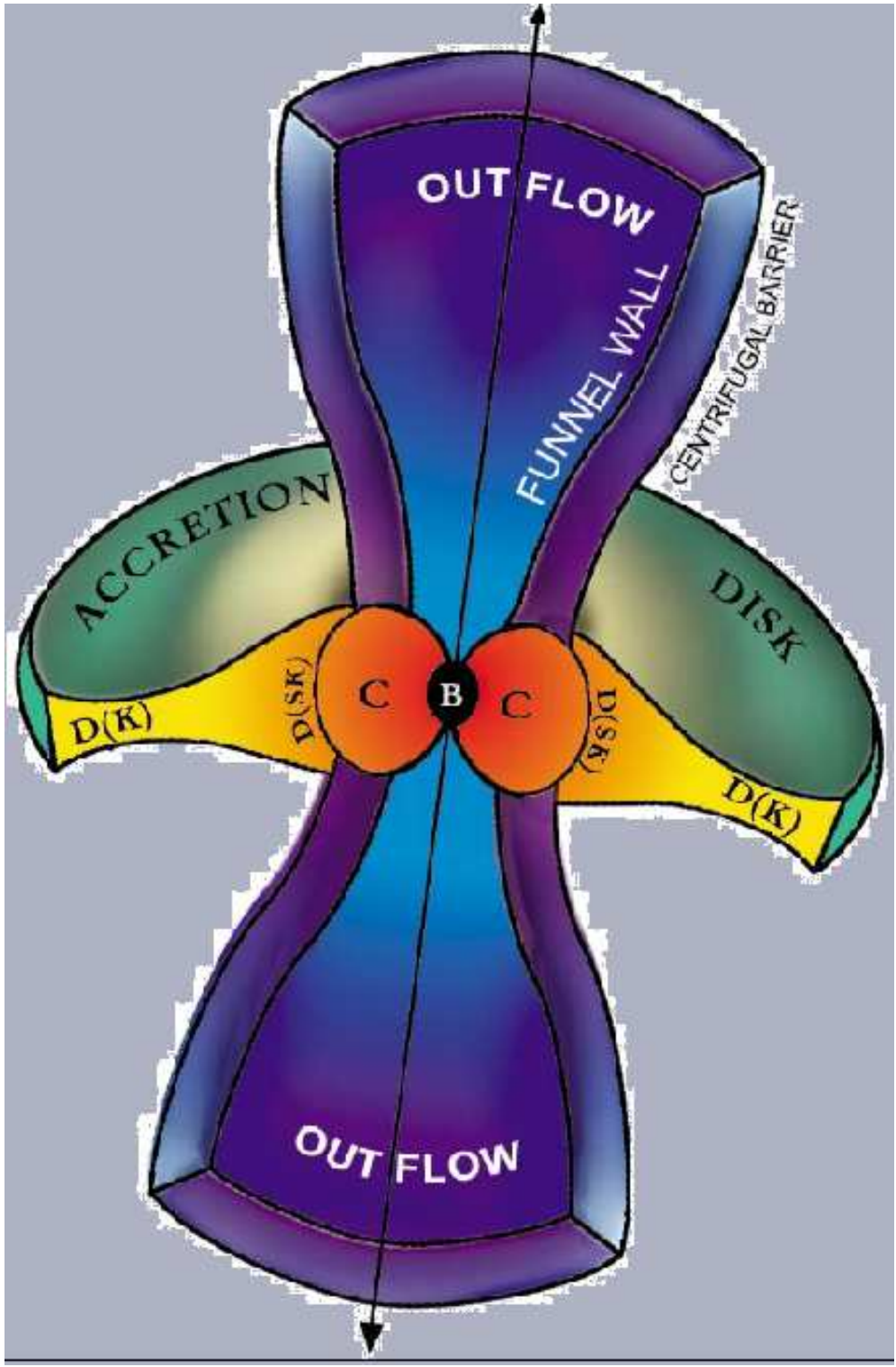}}}
\caption[]{Multicomponent combined flow geometry in 3-Dimension.}
\label{penG}
\end{figure}
In the following figure 
we attempt to illustrate the 
3-D geometry of coupled disk-outflow
system according to our model. {\bf B} is the accreting Schwarzschild
black hole while {\bf C}'s represent the hot and dense CENBOL region.
{\bf D(K)} and {\bf D(SK)} represent the thin Keplarian part and puffed-up
sub-Keplarian part of the advective accretion disk respectively
(measurements not in scale). Due to the axisymmetry assumption in
accretion, two oppositely directed jet are expelled from the close viscinity
of {\bf B}. The inner and the outer surfaces of the outflow are the
funnel wall and centrifugal barrier respectively as explained in Das, 1998 and
in DC99.
\subsection{The Overall Solution Scheme}
Let us suppose that matter first enters through the outer sonic point and
passes through a shock.
At the shock, part of the incoming matter, 
having higher entropy density is likely to return back as winds through
a sonic point, other than the one it just entered. Thus a combination of
topologies, one from the region
of accretion and the other from the wind region is required to obtain
a full solution. In the absence of the shocks,
the flow is likely to bounce back at the pressure maximum of the
inflow and since the outflow would be heated by photons,
and thus have a smaller polytropic constant, the flow would leave the system
through an outer sonic point different from that of the incoming
solution. By simultaneously solving the proper set of equations in appropriate
geometry (see Das, 1998, DC99 and Chapter 2.1 \& 2.2
of Das, 2000a for detail solution scheme and flow geometry),  
we get the {\it combined} flow
topologies (Fig 2. of DC99) where the value of $R_{\dot M}$ along 
with {\it all} other outflow parameters can be exactly computed only from the knowledge
of the {\it minimum number of} accretion parameters namely specific energy
(${\cal E}$),
specific angular momentum ($\lambda$), 
accretion rate (scaled in the unit of Eddington rate)
(${\dot M}_{in}$) and adiabatic indices ($\gamma$) of the 
flow. Also the dependence of  $R_{\dot M}$ on {\it all possible} flow 
parameters has been investigated self-consistently (see Das, 1998 and 
DC99 for detail). 
\section{Outflows from Spherical/Quasi-spherical Accretion}
\subsection{The Outflow Generating Surface}
For this class of accretion, absence  of angular momentum rules out the possibility of
formation of CENBOL.
A novel mechanism is present in the literature 
( Kazanas \&  Ellison, 1986, Protheroe \&  Kazanas, 1983)
where the kinetic energy of spherically
accreting material has been randomized by proposing a steady, 
standing, collisionless,
relativistic hadronic pressure supported
spherically symmetric
shock around a Schwarzschild black hole which produces a nonthermal spectrum
of high energy (relativistic) protons. 
A fraction of the energy flux of infalling matter
is assumed to be converted into radiation at the shock standoff
distance through hadronic ($p -p$) collision and mesonic ($\pi^{\pm},\pi^0$)
decay. Pions generated by this process, decay into relativistic electrons,
neutrinos/antineutrinos and
produces high energy $\gamma$ rays.
These electrons produce the observed non-thermal radiation by Synchrotron
and inverse Compton scattering. 
Shock accelerated relativistic protons are not readily captured by the
black hole  rather
considerable high energy density of these
relativistic protons would be maintained 
to make this shock self-supported (Protheroe \&  Kazanas, 1983).
In this work, we take this
pair-plasma pressure mediated shock surface as the alternative
of the CENBOL from where the outflow could be launched.\\
Here we consider that a Schwarzschild type black hole
quasi-spherically accretes fluid
obeying polytropic equation of state.
We also assume that for our model,
the effective thickness of the shock is small enough compared
to the shock standoff distance. We investigate polytropic as well as isothermal
outflows from polytropic accretion.
\subsection{The Overall Solution Scheme}
The {\it exact} value for the 
shock location and all relevant pre- and post-shock quantities can be
computed in terms of ${\cal E}, {\dot M}_{in}$ and $\gamma$. 
Now using 
the solution scheme as described for the disc-outflow system, the set of equations are 
simultaneously solved in proper geometry to get the combined flow topology
for polytropic and isothermal outflows (see Das 1999, 1999a, 2000 for detail
solution scheme and flow geometry)
 where 
$R_{\dot M}$ along with {\it all other} outflow parameters was calculated and 
its dependence on all flow parameters has been studied.
\section{Some Important Results and Directions for Future Work}
\subsection{Explanation of Quiescent States of X-ray novae}
An interesting situation
arises in our model 
when the polytropic index of the outflow is large and the compression
ratio is also very high. In this case, the flow virtually bounces back
as the wind and the outflow rate can be equal to the inflow rate or even higher,
thereby evacuating the disk completely (see Fig. 6. of Das, 1998).
These cases can cause runaway instabilities by
rapidly evacuating the disk. It is possible that some of the black hole
systems, including that in our own galactic centre, may have undergone
such evacuation phase in the past and gone into quiescent phase. Thus our model 
could explain the quiescent states in X-ray
novae systems like GS2000+25 or  GRS1124-633 etc. (Tanaka, 1995,
Ueda, et al, 1994) and also in some systems with massive black holes,
especially the black hole at our galactic centre.
\subsection{Spectral Properties of Our Galactic Centre}
We suggest that (Das, 1998, Das, 2000a, DC99)
a possible explanation for extreme low luminosity and low radiative efficiency 
(upper limit on the mass accretion rate of $SgrA^{*}$ 
is of the order of $\sim~8\times10^{-5}~{M_{\odot}}~Yr^{-1}$
, Bondi accretion rate on it 
has been approximated as $\sim~3\times10^{-5}~{M_{\odot}}~Yr^{-1}$
(Quarteart, Narayan \& Reid, 1999 and references therein)) could be due to the
presence of profuse mass loss from near viscinity of this source
($SgrA^{*}$). We have obtained that for such a low accretion rate as that
of has been observed for $SgrA^{*}$, the mass outflow rate
is exorbitantly high, almost to the point of evacuating the disk,
which prompted us to strongly speculate that the spectral
properties of our galactic centre could be explained by inclusion of wind
using our model. 
\subsection{Outflow Driven Contamination of Metalicity to the Outer Galaxies}
A number of observational evidences suggest that the fluid accreting onto 
black hole has potential to generate appropriate temperature which 
supports significant  nucleosynthesis to take place
in accretion disks around black holes (Mukhopadhyaya \& Chakrabarti, 
1999 and references
therein).  It is interesting to
investigate whether the fate of the shock induced  nucleosynthesis
generated heavier elements could be predicted by our disk-outflow model.
One of the major speculations of our model (Das, 1998, 
DC99 and \S 5.1 of Chapter 2.2 of  Das, 2000a ) 
that outflows from the hot and dense CENBOL
(where the composition change is much more significant) would carry
away modified compositions and contaminate the atmosphere of the surrounding
stars and galaxies in general. 
Strong indications of disk-evacuation
by wind for some region of parameter space 
suggests  (Das, 1998, DC99)
that overall such contributions to metalicity must not be
neglected. Significant work in this direction is in progress and
is expected to be reported in near future (Das, in preperation).
\acknowledgements
I would like to thank the LOC of this workshop and my special thanks goes to
Prof. A. Castro-Tirado for providing the local and International travel 
hospitality.
{}
\end{article}
\end{document}